\documentclass[manuscript,screen]{acmart}
\AtBeginDocument{%
  \providecommand\BibTeX{{%
    \normalfont B\kern-0.5em{\scshape i\kern-0.25em b}\kern-0.8em\TeX}}}

\setcopyright{acmcopyright}
\copyrightyear{2023}
\acmYear{2023}
\acmBooktitle{CSCW '23 Workshop on Epistemic injustice in online communities, October 2023, Minneapolis, MN.}
\acmConference[CSCW '23 Workshop]{CSCW Workshop on Epistemic injustice in online communities: Unpacking the values of knowledge creation and curation within CSCW applications}{October 14--18,
  2023}{Minneapolis, MN}

\begin{document}

\title[Understanding structured knowledge production]{Understanding Structured Knowledge Production: A Case Study of Wikidata's Representation Injustice}

\author{Jeffrey Jun-jie Ma}
\orcid{0009-0009-0978-0529}
\affiliation{%
  \institution{University of Minnesota}
  \city{Minneapolis}
  \state{MN}
  \postcode{55455}
  \country{USA}}

\author{Charles Chuankai Zhang}
\orcid{0000-0002-1027-9733}
\affiliation{%
  \institution{University of Minnesota}
  \streetaddress{200 Union St SE}
  \city{Minneapolis}
  \state{Minnesota}
  \postcode{55455}
  \country{USA}
  }

\begin{abstract}
Wikidata is a multi-language knowledge base that is being edited and maintained by editors from different language communities. Due to the structured nature of its content, the contributions are in various forms, including manual edit, tool-assisted edits, automated edits, etc, with the majority of edits being the import from wiki-internal or external datasets. Due to the outstanding power of bots and tools reflecting from their large volume of edits, knowledge contributions within Wikidata can easily cause epistemic injustice due to internal and external reasons. In this case study, we compared the coverage and edit history of human pages in two countries. By shedding light on these disparities and offering actionable solutions, our study aims to enhance the fairness and inclusivity of knowledge representation within Wikidata, ultimately contributing to a more equitable and comprehensive global knowledge base.
\end{abstract}



\maketitle

\section{Introduction}
Created in 2012, Wikidata is a structured knowledge base aiming at becoming an open and collaborative database that provides structured data support for other Wikimedia projects. Content in Wikidata is structured and machine-readable, which consists of information of real world entities like books, humans, landmarks, etc. Unlike Wikipedia where different languages have their own edition, Wikidata is multi-language. Each page is accessible to different languages through labels and descriptions featured in the page. The multilingual aspect of Wikidata facilitates contributions from editors speaking different languages, at the same time, the structured nature of Wikidata enables tools and human-created bot accounts to easily perform large volume import and batch edits in a short amount of time. This rapidly expanding nature of Wikidata knowledge makes it prone to biases and injustices. According to previous literature in Wikidata, biases have been found in terms of coverage about gender, occupation, geo-location and race~\cite{shaik2021analyzing, zhang2021quantifying, klein2016monitoring}. The current research primarily focuses on revealing the gap and disparity, but rarely focuses on the source of bias, as well as providing actionable suggestions to address those biases. In this paper, through a case study of comparing human items of two countries, Vietnam and Germany, we propose several reasons that might lead to the existing biases in the knowledge contribution process. The paper structures as follows: First, we present the data and metrics we use to compare human pages in Wikidata. Next, we present the results of our two analyses. Last, we close with discussions about the possible reasons that cause the injustice, actionable solutions, along with presenting future work directions. 

\section{Data}
We chose Germany and Vietnam as subjects based on three primary considerations. Firstly, both nations have comparable population sizes. Secondly, the editors who speak the predominant languages of each country maintain their distinct Wiki communities on Wikidata. In addition, considering the differences between Germany and Vietnam in various aspects, we aimed to investigate how these contrasting contexts might influence online communities' knowledge production.

The dimensions we are comparing are the components each page has along with the attention it received from the community. The components information is collected through Wikidata json dump~\cite{wikidatadump} on the date 2023-07-31. We extracted pages whose \textit{instance of} (P31) properties have the value \textit{human} (Q5) and its \textit{country of citizenship} (P27) properties have the value \textit{Germany} (Q183) or \textit{Vietnam} (Q881). After we collected all the German and Vietnamese human pages, we used Wikidata API~\cite{wikidataapi} to visit its latest 500 edits to collect edit history related information.

\section{Result}
\begin{table}[tbp]
\begin{tabular}{@{}lllll@{}}
\toprule
            & Mean of German pages & Mean of Vietnamese pages & p-value         & effect size \\ \midrule
Label       & 23.31                & 10.76                    & \textless 0.001 & 0.38        \\
Description & 14.45                & 9.02                     & \textless 0.001 & 0.46        \\
Claim       & 20.88                & 13.03                    & \textless 0.001 & 0.63        \\
Sitelink    & 2.47                 & 2.97                     & \textless 0.001 & 0.10        \\ \bottomrule
\end{tabular}
\caption{Welch Two Sample t-test result and effect size for components in German and Vietnamese Wikidata human pages}
\label{tab:t-test}
\end{table}

\subsection{Page content}
The first analysis we did was comparing different components of Wikidata pages between pages in two countries. The components we are comparing are labels, descriptions, claims, and sitelinks. For a single Wikidata page, label is the name that this item is known by, while description is a short sentence or phrase that also serves disambiguate purpose. It's important to note that labels and descriptions are available in multiple languages, making pages more accessible to diverse language communities. Claims document detailed information for each page, ranging from an individual's birthplace to a landmark's geographic coordinates, with more claims indicating a richer information repository within Wikidata. Sitelinks, on the other hand, function as inter-wiki links, connecting the Wikidata page to other Wiki projects. In the dataset we collected, there are 290,750 people who have citizenship of Germany, and there are only 4,744 people who have citizenship of Vietnam. We then performed two-sample t-tests on multiple components, followed by Cohen's D effect size post-hoc test. Table \ref{tab:t-test} shows the result of our analysis. It revealed that there is a statistically significant difference in all components. German pages on average had 13 more labels, 5 more descriptions and 7 more claims compared to Vietnamese pages. While surprisingly, Vietnamese pages had slightly more sitelinks, the difference according to effect size was negligible.

\begin{table}[tbp]
\begin{tabular}{@{}lllll@{}}
\toprule
                              & Mean of German pages & Mean of Vietnamese pages & p-value         & effect size \\ \midrule
Number of total edits         & 73.90                & 44.25                    & \textless 0.001 & 0.54        \\
Number of human edit          & 37.59                & 28,28                    & \textless 0.001 & 0.25        \\
Number of bot edit            & 36.31                & 15.97                    & \textless 0.001 & 0.72        \\
Number of distinct human edit & 16.63                & 11.95                    & \textless 0.001 & 0.42        \\
Number of distinct bot edit   & 14.60                & 8.34                     & \textless 0.001 & 0.71        \\ \bottomrule
\end{tabular}
\caption{Welch Two Sample t-test result and effect size for German and Vietnamese Wikidata human pages edit history}
\label{tab:edit-history}
\end{table}

The second analysis focused on the edit history of Wikidata items. Through the API, we were able to extract the edit history of Wikidata items. The feature we extracted is the user's attention. Specifically, we extracted the username of each revision and checked if the user was bot or not. Then we quantified the attention metric into five features: Number of total edits, number of human edits, number of bot edits, and number of distinct bot and human edits. Table \ref{tab:edit-history} shows the comparison of these five features within two countries. We can observe that in all the five features the result is significant and in terms of bot activity and total activity, the effect size is beyond medium threshold (0.5). 

\section{Discussion}
\subsection{Reflected biases in different dimensions}
In our analysis, we conducted a comparative examination of various attributes between human pages related to Vietnam and those related to Germany within Wikidata. The findings revealed substantial disparities. It was evident that not only are there significantly fewer individuals from Vietnam represented in Wikidata, but the average content on their respective pages was also lower when compared to individuals from Germany. This discrepancy extended to the quantity of labels, descriptions, and claims, indicating that Vietnam-related pages are accessible in fewer languages and contain less information overall. The results from the comparison of attention features also revealed a massive difference in terms of activities of Wikidata pages. Vietnamese human pages have fewer human and bot edits, with bot edits having a larger disparity. This indicates that not only are there fewer human editors dedicated to enhancing the quality of Vietnam-related pages, but it also indicates that bot developers may have focused their efforts more towards importing German-related items or maybe drawing from biased data sources when contributing to Wikidata.

\subsection{Source of injustice}
Based on our analysis in API as well as our subsidiary research, we conclude three reasons that cause this bias: difference in existing within-community knowledge, difference in editor base, and difference in bot power. From the stats collected in Wikimedia, German Wikipedia features 2,839,259 articles with 17,949 active users, while Vietnamese Wikipedia only has 1,288,010 articles with 1,935 users. Since Wikipedia serves as a major data source of Wikidata, the vast difference in Wikipedia language communities creates an unjust base for Wikidata items to develop on. In terms of Wikidata users, recent language proficiency data shows that 3,419 Wikidata editors declared proficiency in German while only 90 editors declared proficiency in Vietnamese. Although there's only a small batch of editors declaring language in Wikidata, the difference in declared proficiency also indicates the difference in terms of editor base in Wikidata. Lastly, the bot activity difference indicates that tool editors and bots creators edit more in German than in Vietnamese. Considering the power of automated editing in creating structured information, the difference in bot attention could create a huge coverage gap in terms of knowledge contribution.

\subsection{Mitigating the issue: tool developer awareness and the introduction of LLMs}
This analysis could be generalized to any two language communities, and it could be foreseen that the community with more editors and bot developers would have a larger volume of content in this multi-language platform. Thus it is very important for the Wikidata community to come up with solutions that aim at increasing the awareness of justice for Wikidata editors. Especially for those editors that use tools or bots to perform large batch edits, considering the large volume of contribution~\cite{steiner2014bots, piscopo2017makes} made in Wikidata, it would be useful to either interest them towards less represented fields or towards more diverse data sources. 

Furthermore, as large language models (LLM) are more widely used in our lives, considering the accuracy and power of recently released LLM tools such as GPT-4, the usage of LLMs and its power in accurate translation~\cite{jiao2023chatgpt} could also be considered when trying to improve editing in certain components like labels and descriptions. With thorough testing and validation procedures, semi-automated tools could be developed to assist editors with their editing. For example, when creating an item, with the LLM's assistance, it could generate multi-language labels and descriptions with the editor's manual edit of monolingual label and description edit. Yet what the influence of deploying these tools on a large scale brings to the edit process, quality control, and vandalism detection remains to be further investigated. 

\section{Conclusion}
In this paper, through a case study, we examined the difference of Wikidata pages from two dimensions: quality and attention. We identified a large difference and presented several preliminary assumptions of the reason that causes those disparities. To summarize, the structured nature of Wikidata followed by its largely automated knowledge contribution process makes it prone to biases and injustice that exists in third-party data sources. Future work shall focus on further decomposing the cause of injustice, and come up with tools and framework to mitigate the issue.

\bibliographystyle{ACM-Reference-Format}
\bibliography{sample-base}
\end{document}